\documentclass[review, sort&compress, 5p]{elsarticle}
\pdfoutput=1
\usepackage{array}
\usepackage{hepunits}
\usepackage{SIunits}
\usepackage{hepnames}
\usepackage{float}
\usepackage{dblfloatfix}
\usepackage{booktabs}
\usepackage{perpage}
\usepackage{mdwmath}
\usepackage{mdwtab}
\usepackage{hyphenat}
\usepackage{natbib}
\usepackage[font=small,labelfont=bf]{caption}
\usepackage[colorlinks=true, urlcolor=blue, linkcolor=blue, citecolor=blue, raiselinks=true, hyperfootnotes=true]{hyperref}

\usepackage{lineno}

\begin{document}

\begin{frontmatter}

\title{Timewalk correction for the Timepix3 chip obtained with real particle data}


\author[mymainaddress]{S. Tsigaridas\corref{cor1}}
\author[secondaddress]{M.~v.~Beuzekom}
\author[secondaddress,thirdaddress]{H.~v.~d. Graaf}
\author[secondaddress]{F.~Hartjes}
\author[secondaddress]{K.~Heijhoff}
\author[fourthaddress]{N.~P.~Hessey}
\author[secondaddress]{P.~J.~de Jong}
\author[secondaddress,thirdaddress]{ and V.~Prodanovic}

\cortext[cor1]{Corresponding author. Tel.: +33 (0)4 76 88 25  42.\\ E-mail address: \href{mailto:stergios.tsigaridas@esrf.fr}{stergios.tsigaridas@esrf.fr (S. Tsigaridas)}\\}
\address[mymainaddress]{European Synchrotron Radiation Facility, 71 Avenue des Martyrs, 38000 Grenoble, France}
\address[secondaddress]{Nikhef, Science Park 105, 1098 XG Amsterdam, The Netherlands}
\address[thirdaddress]{Delft University of Technology, 2628 CT Delft, The Netherlands}
\address[fourthaddress]{TRIUMF, 4004 Wesbrook Mall, Vancouver, BC, Canada V6T 2A3}

\begin{abstract}

In this work we have developed a prototype gaseous pixel detector by combining a micromegas grid with a Timepix3 chip for the readout. The micromegas foil supported by a matrix of pillars about 50~$\micron$ high was manually placed on top the chip. By placing a cathode foil above the chip an ionisation detector was created with a drift gap of 13.5~$\mm$. The Timepix3 chip, thanks to the simultaneous measurement of the time-of-arrival (ToA) and charge via time-over-threshold (ToT) allows corrections to remaining timewalk effects, improving further the position resolution along the drift direction. We present the timewalk correction for Timepix3 chip obtained with real data from a particle beam and its impact on the tracking performance. The results obtained show a significant improvement on the position resolution for single-hits and tracks along the drift direction compared to previous experiments. 

\end{abstract}

\begin{keyword}
Micro-pattern gaseous detector, Gaseous pixel detector, Timepix3, GridPix, GasPix, Timewalk, Tracking.
\end{keyword}

\end{frontmatter}

\section{Introduction}

The gaseous pixel detector GridPix is a novel detector technology for particle tracking which combines a micromegas amplification grid with a pixel chip for the readout. The chip is covered with a thin resistive layer to prevent potential damage from discharges. Subsequently, the grid added by photolithography or manually placed directly above the chip to create an amplification region. To create a drift field, a cathode electrode is added above the grid. The whole structure is placed in a gas filled chamber. With this technology we are able to measure in 3D the creation position of individual ionisation electrons released by traversing particles and thus to reconstruct their trajectory.   

In a previous work \cite{ref_Gossipo2} we measured the tracking performance of a gaseous pixel detector using the Gossipo2 chip and  featuring a high-resolution TDC (1.8~ns) per pixel. The performance of this prototype chip technology led to a significant improvement to the position resolution in the drift direction. Its successor, the Timepix3 chip \cite{ref_Timepix3}, implements a TDC per pixel with a better resolution of 1.56~ns, which enhances the performance. In addition, the simultaneous measurement of  charge via time-over-threshold (ToT) and drift time via time-of-arrival (ToA) allows corrections  to remaining timewalk effects, improving further the resolution. These facts encourage the realisation of gaseous pixel detectors based on the Timepix3 chip. Therefore we assembled a gaseous pixel detector based on the Timepix3 chip that we called GasPix and installed it in a particle beam. Our aim was to study the correction for the remaining timewalk effects and measure the impact on the tracking performance.


\section{Detector}
\label{lab_DetectorDescription}

\begin{figure}[!h]
\centering
\includegraphics[width=1.\linewidth]{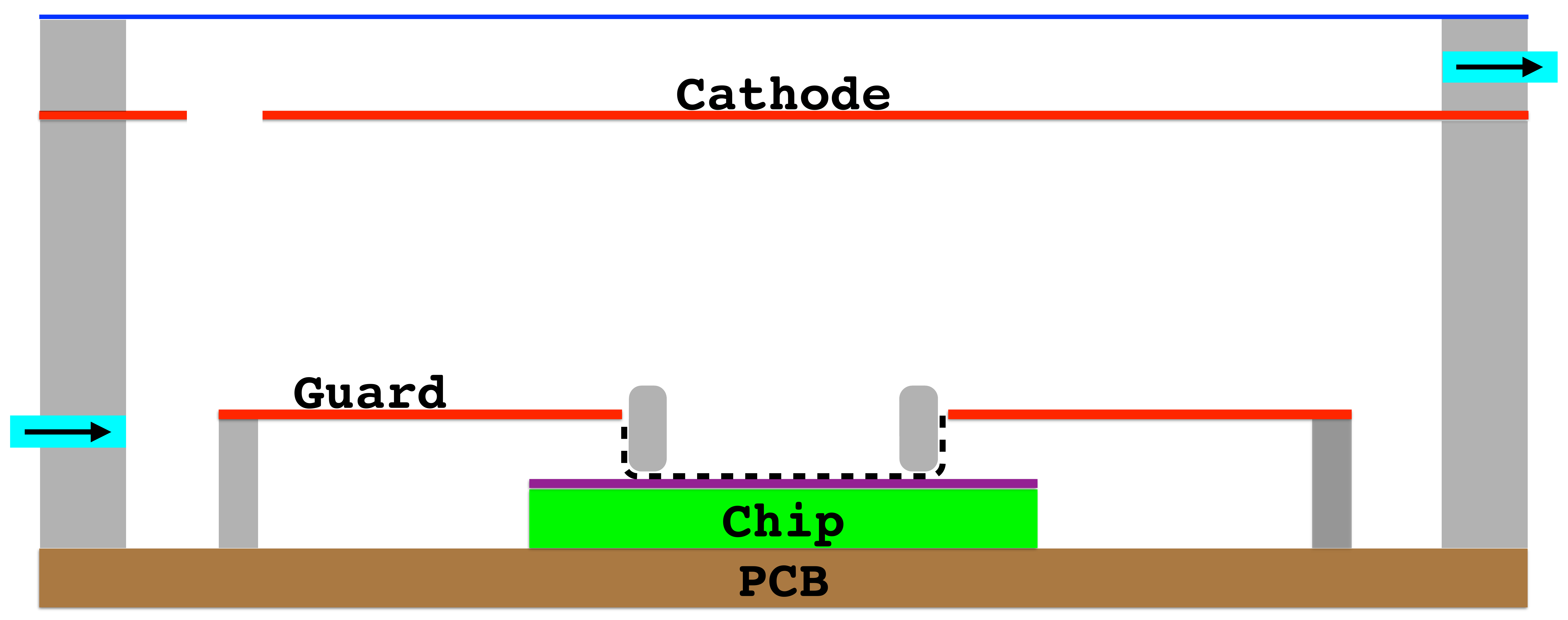}
\caption{A cross section of the detector (not in scale). The grid (dashed line) is glued to a frame and then mounted on the chip. The frame and the cage are made from the same material. }
\label{fig_detector}
\end{figure}

The cross section of the detector is illustrated in Fig.~\ref{fig_detector}. Beginning from a bare Timepix3 chip, we covered its sensitive area with a 4~$\micron$ thick SiRN protection layer\footnote{The deposition of the layer was performed in  the Else Kooi Laboratory (EKL) at Delft University of Technology resulting in a layer with a resistivity~$<$~10$^{11}$~$\ohm\cdot\cm$.} against discharges. After the deposition of the protection layer, the chip is glued with a thin double-sticky tape and wire bonded on a specially designed chip carrier board known as Spidr-board. 

A frame glued on the Spidr-board holds the guard electrode above the Timepix3 chip such that its laser-cut window is aligned to the chip. The micromegas grid was glued onto a frame of Ertalyte fitting just into the aperture of the guard. Subsequently the frame was attached with a minor amount of glue to the chip.

\begin{figure}[!h]
\centering
\includegraphics[width=1.\linewidth]{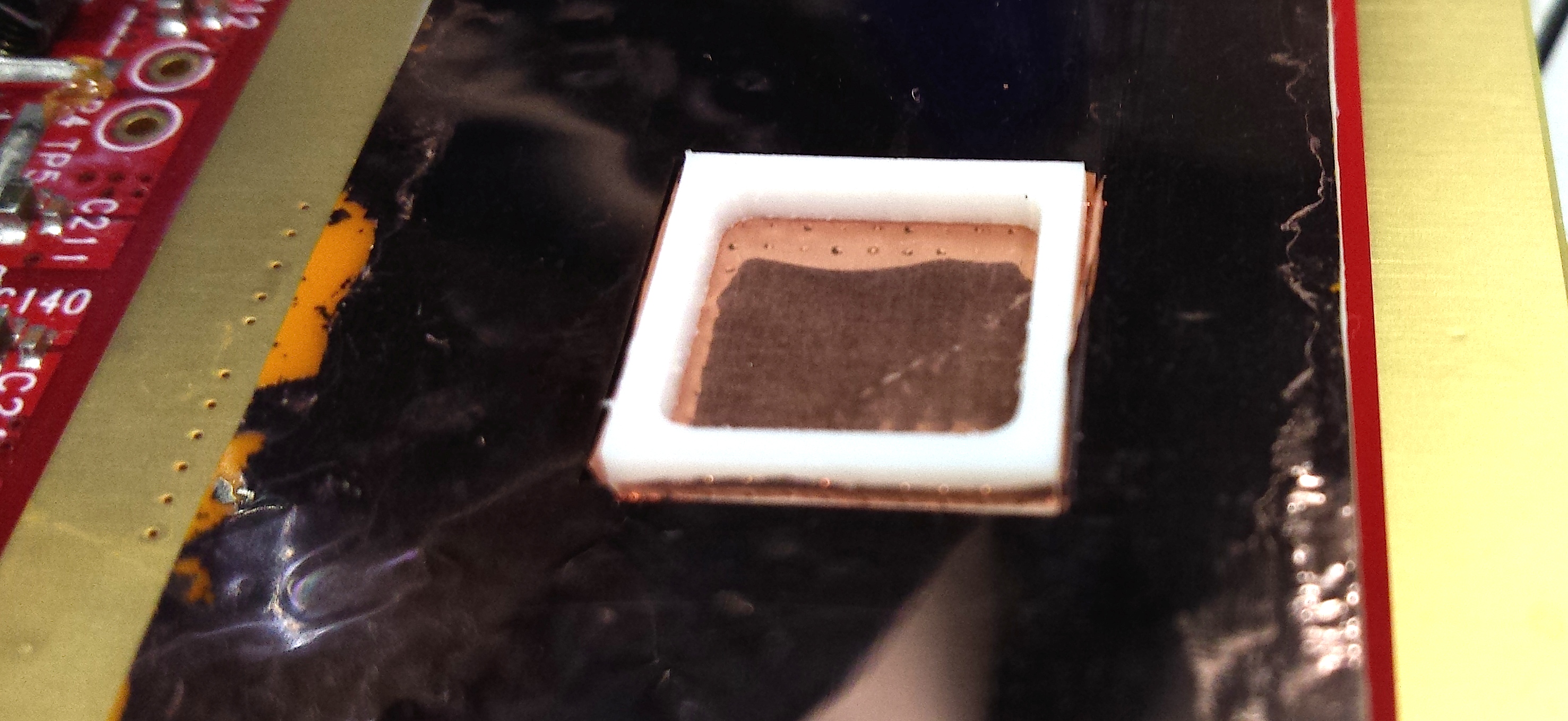}
\caption{Photograph taken after the assembly of the grid structure, bonded to the white frame, on top of the chip.}
\label{fig_final}
\end{figure}

The micromegas grid was fabricated at CERN on a 5~$\micron$ thick copper plate on a kapton backing by chemical processing a structure consisting of a copper plate with small holes supported by tiny Kapton pillars about 50~$\micron$ high. For more details, see \cite{ref_etching}. The grid-holes had a pitch of 60~$\micron$ and a diameter of 35~$\micron$. The pillars with a spacing of 1~$\mm$ and the electrostatic force of the grid voltage provide a constant distance between  grid and chip. Unfortunately the grid-hole pitch is bigger than the pixel pitch of 55~$\micron$. Due to this mismatch a fraction of the pixels are reasonably aligned to a grid hole. Care was taken to make the grid parallel  to the pixel pattern of the chip. A small amount of araldite on the outer side of every corner of the grid-frame keeps everything in place. The result is shown in Fig.~\ref{fig_final}. 

The grid is biased by a direct contact to the guard. A small extension of the grid at one of its corners, is put in contact with conductive glue. We were aware that this geometry will cause electric field deformations. However, since no other solution was available at that time, we decided to proceed with this solution and then correct offline for remaining effects. 

The detector was completed by adding the cathode and enclosing everything into a gas-tight chamber yielding to a drift gap of 13.5~$\mm$. The cathode is a 5~$\micron$ thick copper foil to allow the passage of low energetic particles from a source.  It is supported by 2~$\mm$ thick plate of Ertalyte leaving open a 15 mm wide circular hole above the chip. This plate is mounted on a precision machined frame. The base of the frame has an O-ring seal to the Spidr-board.

\begin{figure}[!h]
\centering
\includegraphics[width=1.\linewidth]{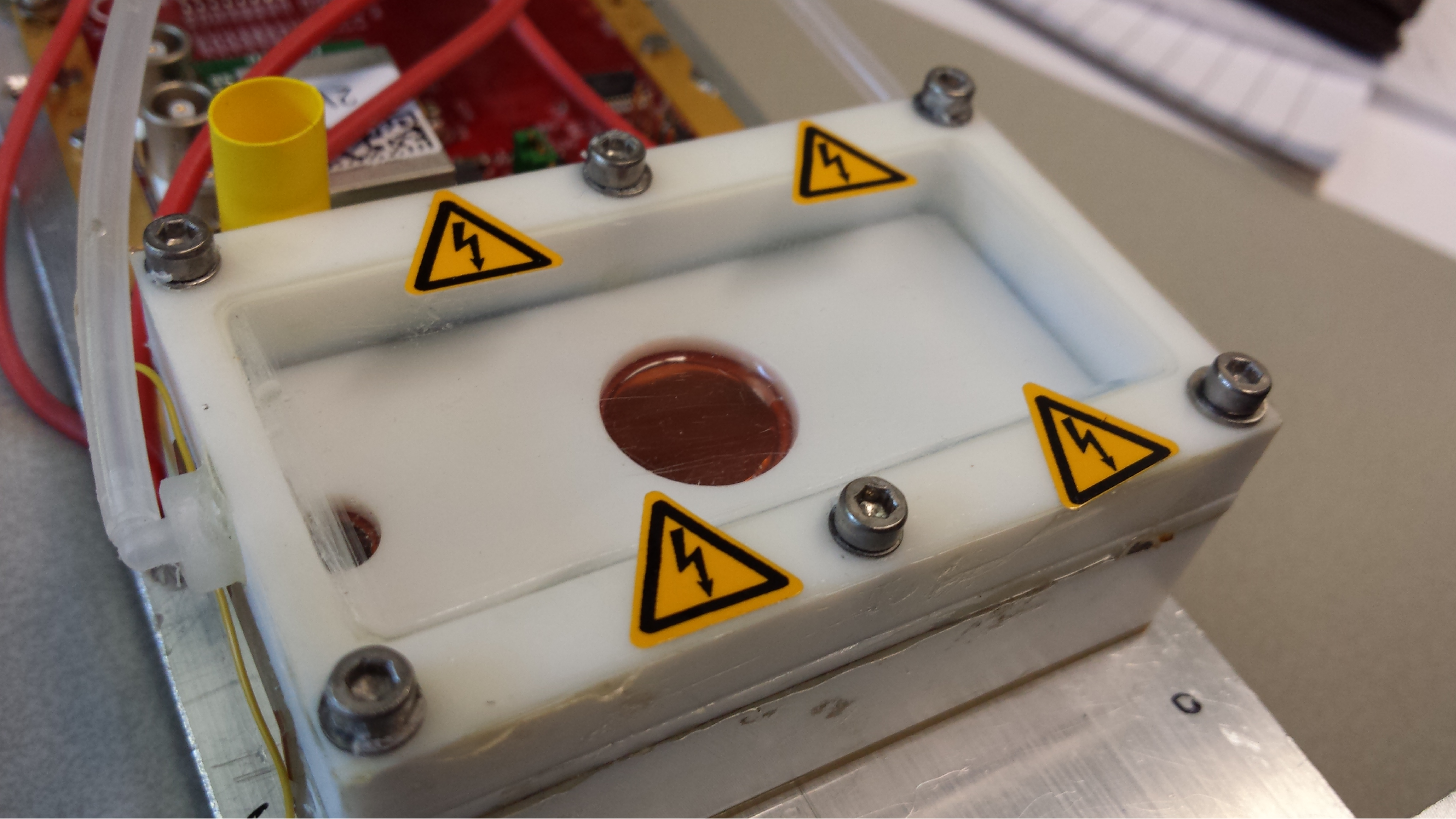}
\caption{The gas enclosure. The top lid (transparent) holds the gas. The cathode is visible in the centre.}
\label{fig_field_cage}
\end{figure}

A rectangular box with a clear mylar window is placed on top of the cathode and bolted to the Spidr-board, making a gas tight seal. A hole in the cathode support allows the supply of chamber gas, preventing the displacement of the cathode foil due to the chamber overpressure. The gas is exchanged through a hole in the gas chamber walls. The total volume of the gas chamber is about 45~$\cm^3$. Fig.~\ref{fig_field_cage} gives a close overview of the chamber.

\section{Experimental testbeam setup}
\label{lab_Testbeam}
 
For about a week in the end of August 2015, we collected testbeam data at the H8 line of the Super Proton Synchrotron (SPS) at CERN. The H8 line delivers usually a high intensity beam of  180~$\GeVoverc$ hadrons (60\%~protons/40\%~pions). Within a typical beam cycle of 36~$\second$, two spills are present with a duration of 4.8~$\second$ each. The beam is quite focused with a maximum rate of about 1$\cdot 10^6$ particles~/~spill per~$\centi\meter\squared$. 

\begin{figure*}
\centering
\includegraphics[width=.8\linewidth]{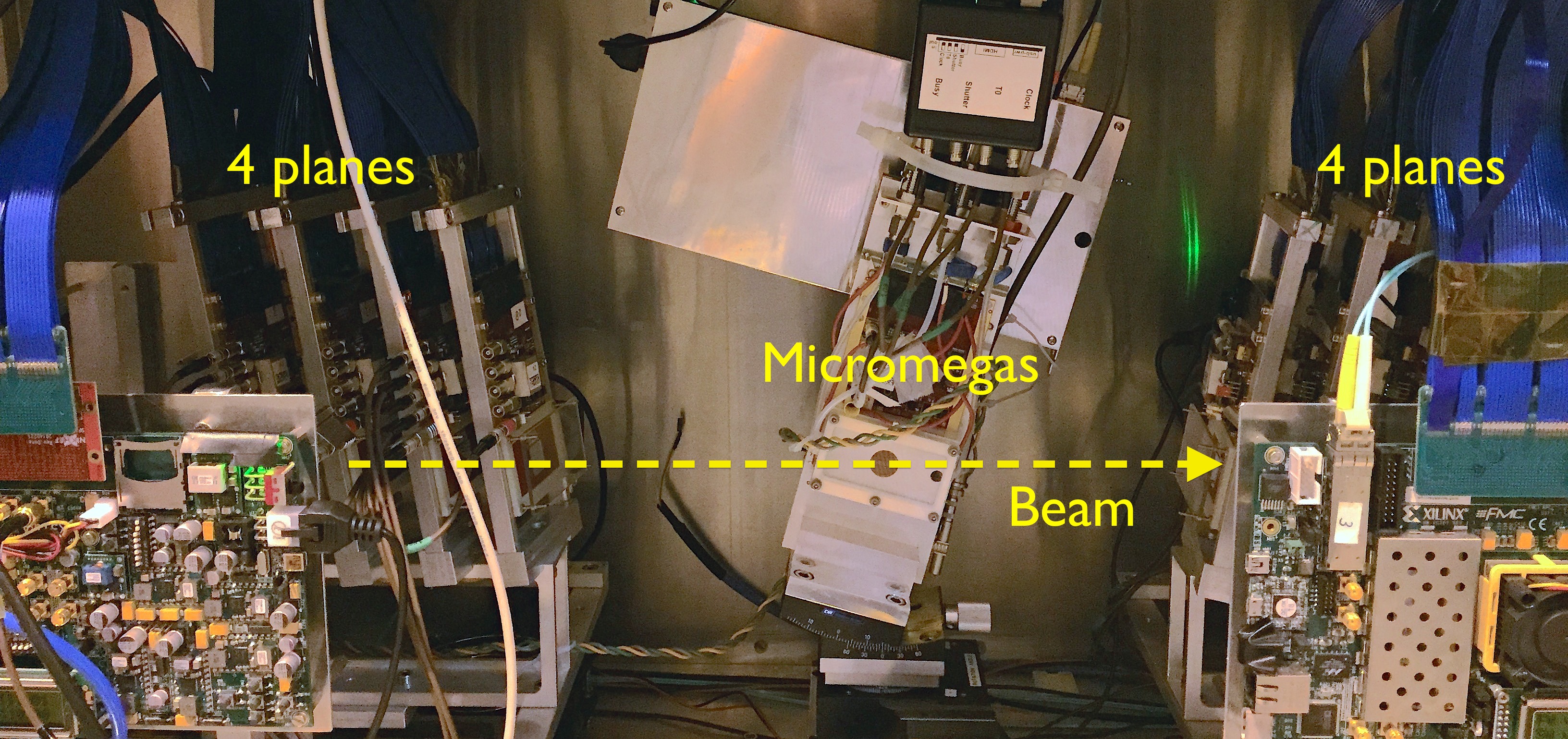}
\caption{Photograph of the experimental setup taken during the testbeam.}
\label{fig_tel} 
\end{figure*}

To have an external track reference a Timepix3 silicon telescope was used, offered by the LHCb VELO upgrade group. The telescope consists of two arms, each of them carrying 4 stations. Each telescope station is a Timepix3 hybrid pixel detector. Between the arms there was enough space to mount the device under test. The telescope is able to reconstruct extremely precise tracks with a pointing resolution of $\sim$2~$\micron$. For triggering purposes, before the first and after the last station, there are two scintillators covering an area of 1~$\cm^2$ each.  Fig.~\ref{fig_tel} shows the experimental setup used. 

The readout of the GasPix and the telescope is done by the SPIDR system for the Timepix3 chip, \cite{ref_spidr}. This was extremely convenient since we merge the data of the device under test in the same data stream. Thanks to the SPIDR's high readout speed every track is recorded.

The chamber was flushed with a CO$_2$/DME (50/50\%) gas premixed at Nikhef. While running, a moderate flow of 5~$\cm^3\per\min$ was used. The input and exhaust flow are controlled and measured respectively by mass flow technology running under LabView. 

For the results presented in this work the grid voltage was set at -550~$\volt$ enabling a gas gain of 1500-2000\footnote{The gain was measured in the lab using a $^\mathrm{55}$Fe source, \cite{ref_Thesis}.}. The cathode voltage was set to yield  a drift field of 1~$\kilo\volt\per\cm$. The detector was mounted on a goniometer adjusted at a tilt angle of 15$^\circ$ with the horizontal axis. The goniometer was mounted on a remotely controlled rotation stage allowing rotations about  a vertical axis and a translation stage which allows translations on a plane perpendicular to the beam. The rotation angle was set at 20~$^\circ$ and the detector was operated at the minimum useful threshold of 500~$\Pelectron$.

\section{Data reconstruction}
\label{lab_reco}

\subsection{Telescope data}
\label{lab_TelAnalysis}

The raw telescope data were analysed using the standard LHCb software called  \textsc{~Kepler} which is a part of the \textsc{Gaudi}\footnote{\href{http://gaudi.web.cern.ch/gaudi/}{http://gaudi.web.cern.ch/gaudi/}} framework. This finds the trajectories of the beam particles and stores these as tracks. The signature of a track in each telescope station is a cluster of pixels hit and its cluster time. The mean cluster time of all the stations is the track timestamp associated to each track. The \textsc{Kepler} program is used for the alignment and tracking for the telescope. The input to the  \textsc{~Kepler}  program is an ASCII file which contains the nominal alignment constants. We perform the alignment of each station using the Millipede\footnote{\href{http://www.desy.de/~blobel/mptalks.html}{http://www.desy.de/~blobel/mptalks.html}} method. Then, by using the updated alignment constants the tracking is performed. The typical minimum requirement for a track is the presence of a cluster in each station. 

Every track is projected to the $xz$ and the $yz$ plane and fitted with a straight line minimising the $\chi^2$. The final output is a ROOT\footnote{\href{https://root.cern.ch/}{https://root.cern.ch/}} file which contains all the information related to the tracks, i.e the track times, the slopes and offsets for the $xz$ and the $yz$ plane along with the covariance matrix and the $\chi^2$ values. 

\subsection{GasPix data}

The raw GasPix data is processed into events where each hit is associated to a scintillator trigger timestamp. All the hits within a time window around a trigger time\hyp{}stamp are assigned to an event. The output is an ASCII file which contains the event number, the trigger time and the associated hit information; namely the column and row number, the hit arrival time and the time over threshold. Then the data are decoded and stored in a ROOT tree. This event data is then paired with a telescope track using the two timestamps, and merged into a single file for further analysis. 

\section{Analysis}
\label{lab_Analysis}

During the analysis, the raw hit information is calibrated into a 3D hit position $(x_h, y_h,z_h)$ in mm in the GasPix frame. As discussed we expect drift field deformations which we can correct offline thanks to the very precise telescope track. An iterative alignment procedure gives corrections to hit positions and transformations of telescope tracks into the GasPix frame. Events with telescope tracks in a good fiducial volume and away from the edges are selected. Displacement of GasPix hit positions from the telescope tracks give residuals plots, which can be used to correct for things like drift field variations. With less distortions in hit positions, the alignment coefficients can be improved.

The procedure is very iterative: better alignment gives better hit position correction, which in turn gives better residuals for a better alignment. After all hit position corrections and alignment constants are stable, the residuals are plotted against ToT. This is used to optimise the timewalk correction. After applying all the corrections, including for timewalk, local track fits are made to the GasPix hit positions. The GasPix track resolution is extracted from the covariance matrix of the fit. 

In the following section we present the formulae used in the analysis. For a detailed description of the analysis procedure the reader is referred to \cite{ref_Thesis}.

\subsection{Hit position calibration}

When the grid and the pixel matrix, have the same pitch and are well aligned, the grid-hole position is identical to the pixel position; the pixel position would be given by the following equations:
\begin{align}
	x_p &= d_p \cdot(n_{c} - x_0) \,, \label{eq1.1}\\
	y_p &= d_p \cdot(n_{r} - y_0)  \,. \label{eq1.2}\
\end{align}     

In eq.~\ref{eq1.1},\ref{eq1.2}  $n_c$, $n_r$ are the column and row numbers in pixel units which vary from 0 to 255 while $d_p = 55~\micron$ is the pixel pitch. These are shifted by $(x_0, y_0) = (127.5, 127.5)$ so that the origin of the local frame is the centre of the chip. In case of a mismatch though, the true hit position is the nearest grid-hole position. This is given as follows:
\begin{align}
	x_h &= \mathrm{nint} \left ( \frac{x_p - x_c}{d_g} \right ) \cdot d_g  \,, \label{eq1.3}\\
	y_h &= \mathrm{nint} \left ( \frac{y_p - y_c}{d_g} \right ) \cdot d_g  \,. \label{eq1.4}
\end{align}     

In eq.~\ref{eq1.3},\ref{eq1.4} $d_g = 60~\micron$ is the grid-hole pitch and  $(x_c, y_c) = (30, 30)~\micron$ are the offsets in order to match the centre of the grid with the origin. The method also allows offsets in the rotation angle. Special care was taken during the mounting of the grid so that the angle was consistent with zero. 
   
The reconstruction of $z_h$ is based on the drift time given by 
\begin{align}
	t_{d} & = \left( t_{ar} - t_{tr} + t_{phase} \right) - t_0 \,, \label{eq1.5}\\
	z_h &= v_d \cdot t_d \, \label{eq1.6}\,.
\end{align}   

In eq.~\ref{eq1.5} $t_{ar}$ is the hit arrival time and $t_{tr}$ is the trigger timestamp of the event and $t_{phase}$ is an column-dependent shift in the arrival time due to the fact that in Timepix3 the clock is not generated synchronously for all the columns. The $t_{ar}$ value has to be shifted by $t_0$ so that the earliest hits are produced at  $z = 0$. Then, the $z_h$ in mm is the product of the drift velocity $v_d$ and the drift time $t_d$. Below we modify eq.~\ref{eq1.6} to account for electric field deformations.

\subsection{Electric field deformations}

For the current detector it was well known in advance that the drift field would not be perfect. One reason is that the grid and the guard are at different heights but the same potential which leads to deformed electric field on the edges of the grid. This is made even worse by possible charging up the frame above the grid. Distortions in the grid and the cathode also alter the electric field. 

The central part of the chip is expected to have an electric field which is highly uniform while the electric field on the edges is distorted. For the analysis we select tracks away from the walls to avoid the known distortions. Still, due to the remaining distortions the drift velocity is not constant across the remaining detector volume. Therefore, the assumption of a constant drift velocity for the z-position reconstruction is not useful. But for the calibration of the distortions we can make use of the reconstructed tracks in the telescope. This enables the calculation of a correction for the $z_h$ which is independent of the drift velocity $v_d$.

\begin{figure}[!ht]
\centering
\includegraphics[width=0.9\linewidth]{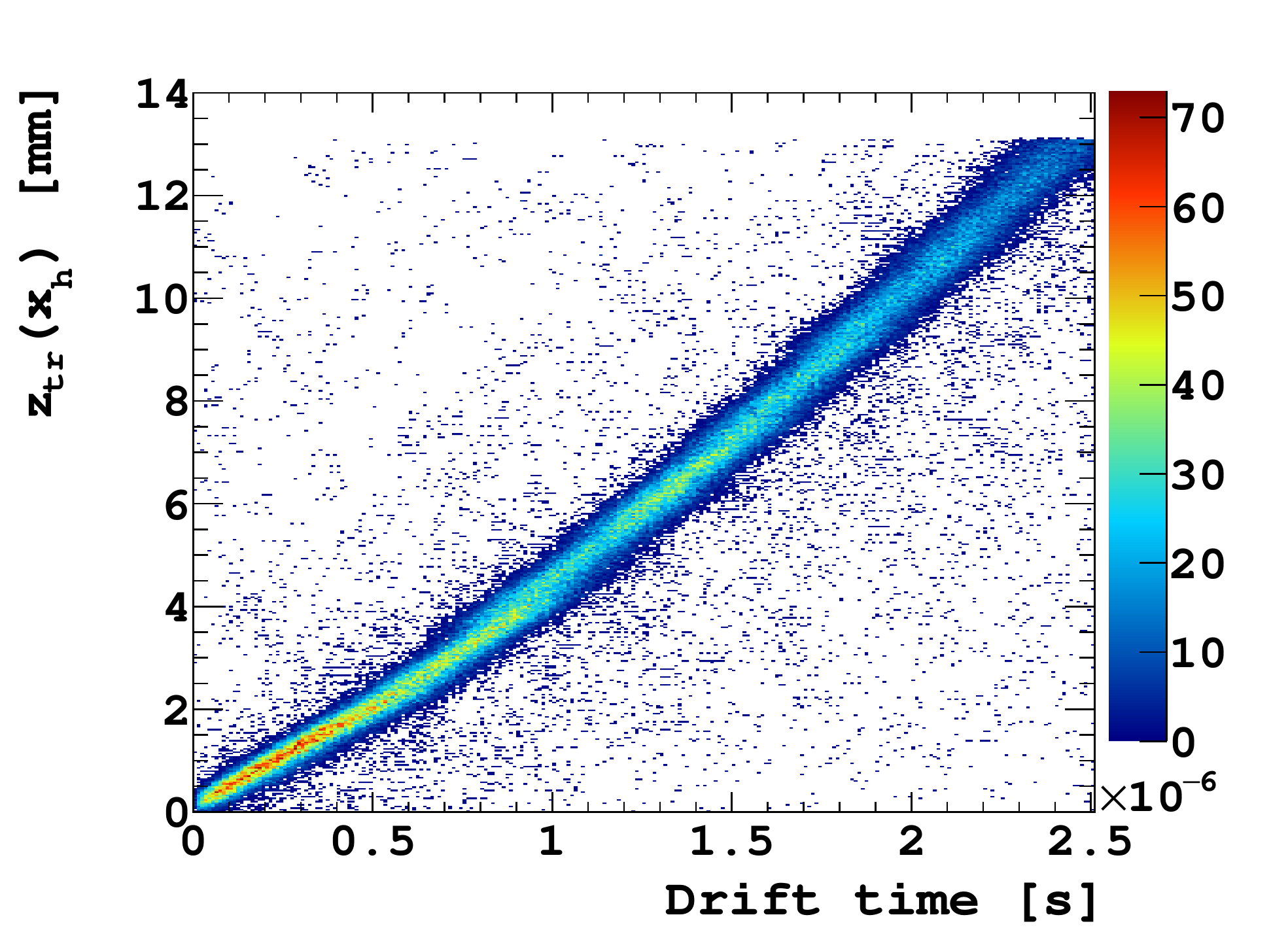}
\caption{The 2D histogram of the telescope predition against the drift time. For the ideal detector without electric field distortions the relation is linear. }
\label{fig_zcorr} 
\end{figure}

The telescope track is used to calculate $z_{tr}$ which is the prediction for the $z_h$ position of each hit. This is done separately for the $xz$ and the $yz$ projection of the telescope track (transformed in the local frame) and gives the nearest point to $x_h$ and $y_h$. Then $z_{tr}$ is plotted against the corresponding drift time of each hit, Fig.~\ref{fig_zcorr}. Each slice of the drift time is projected to the $y$-axis and fitted with a gaussian. The profile histogram of the fitted means is fitted with a Chebyshev polynomial (eq.~\ref{eq_zcorr}) to give $z^{'}_h$ as a function of drift time:
\begin{equation}
	z_{h}^{'} = z_{tr} = \sum \mathrm{Cheb}\,( t_d)\,.
	\label{eq_zcorr}
\end{equation}

The drift velocity is then given by the derivative of the eq.~\ref{eq_zcorr}:  
\begin{equation}
	v_{d}^{'} = \frac{\mathrm{d} }{\mathrm{d} t_d} 
 \sum \mathrm{Cheb}\,( t_d)\,.
	\label{eq_vdcorr}
\end{equation}

\subsection{Timewalk}

\begin{figure}[!h]
\centering
\includegraphics[width=\linewidth]{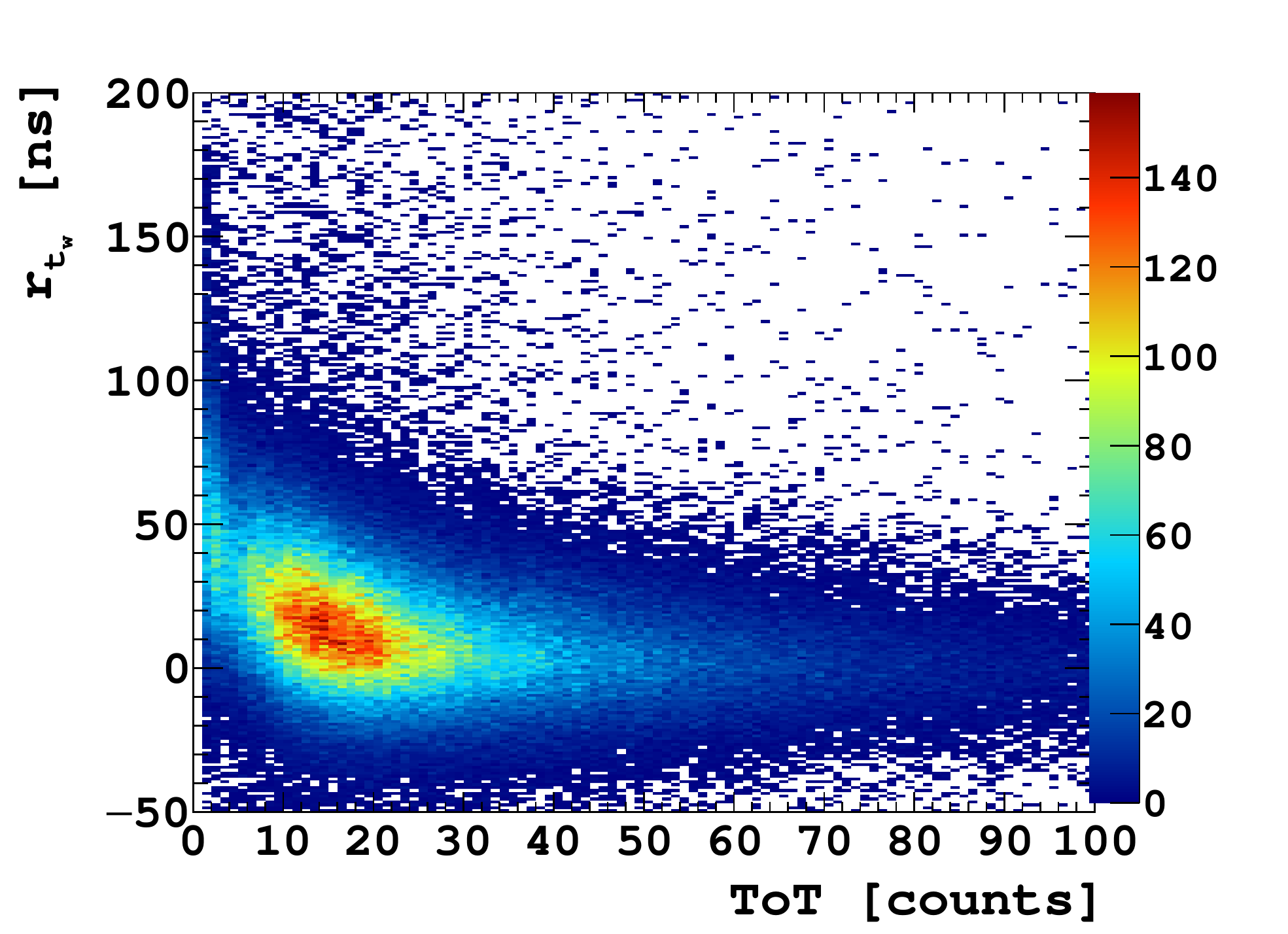}
\caption{The 2D histogram of the residual $r_{t_w}$ against the ToT that is used for the timewalk correction. For low ToT counts, the residual distribution has a tail to the positive direction due to timewalk.}
\label{fig_tw} 
\end{figure}

The successful calibration of the hit positions and the alignment of the detector to the telescope enable the correction for remaining timewalk effects. The correction is obtained from a 2D histogram of the residual $r_{t_w}$, given by eq.~\ref{eq_tw}. This residual is the closest distance from the hit to the fitted telescope track translated in time and is plotted against ToT in Fig.~\ref{fig_tw}.  
\begin{equation}
	r_{t_w} = \frac{r_{xz}(z^{'}_{h})}{v^{'}_d} = \frac{ \big( x_h - x_{tr}(z^{'}_h) \big)_\perp}{v^{'}_{d}}  \,.  \label{eq_tw}
\end{equation}

Each ToT slice is then fitted with a gaussian. The fitted mean value of each slice is the timewalk and is used to calculate $z^{''}_h$ which is the $z$ position corrected for timewalk, eq.~\ref{eq_zpp}. The fitted sigma value includes the contributions of diffusion and timewalk. By subtracting the contribution of the diffusion we extract the contribution of the timewalk to the error on $z^{''}_h$. The correction is applied directly on the raw data i.e on the ToT and so there is no need to convert the ToT to charge,   
\begin{equation}
	z^{''}_{h} = z^{'}_{h} - v^{'}_{d}\cdot t_w(\mathrm{ToT}) \,.
	\label{eq_zpp}
\end{equation}

\subsection{Track fitting}
 
Each event consists of 3D hits with coordinates given by ($x_h$,~$y_h$,~$z_{h}^{''}$) where $z_{h}^{''}$ is the height of the ionisation cluster corrected for field deformations and timewalk, given by eq.~\ref{eq_zpp}. A straight line is fitted through the projections in $xz$, $yz$ planes. For the fitting we use the same method, as in \cite{ref_Gossipo2}.  The errors assigned to $x_h$ and $y_h$ positions are identical due to the symmetry of the detector and given by

\begin{equation}
\sigma^2_{x_h} = \sigma^2_{y_h} = \frac{d^2_g}{12} + D^2_T\,z \,, \label{eq_errorXY}
\end{equation}

while for the $z^{''}_h$ the error is given by

\begin{equation}
\sigma^2_{z^{''}_h} = \frac{(\tau_f\,v_d')^2}{12} + D^2_L\,z + (v_d' \, \sigma_{t_w}(\mathrm{ToT}))^2.            \label{eq_errorZ}
\end{equation}

In eq.~\ref{eq_errorXY}, \ref{eq_errorZ} $d_g$ is the grid-hole pitch, $D_T$ and $D_L$ are the transverse and longitudinal diffusion coefficients, $\tau_f = 1.56~\ns$ is the TDC resolution for the Timepix3 chip, $v^{'}_d$ is given by eq.~\ref{eq_vdcorr} and $\sigma_{t_w}(\mathrm{ToT})$ is extracted from the timewalk correction.     

\begin{figure}[!h] 
\centering
\includegraphics[width=0.9\linewidth]{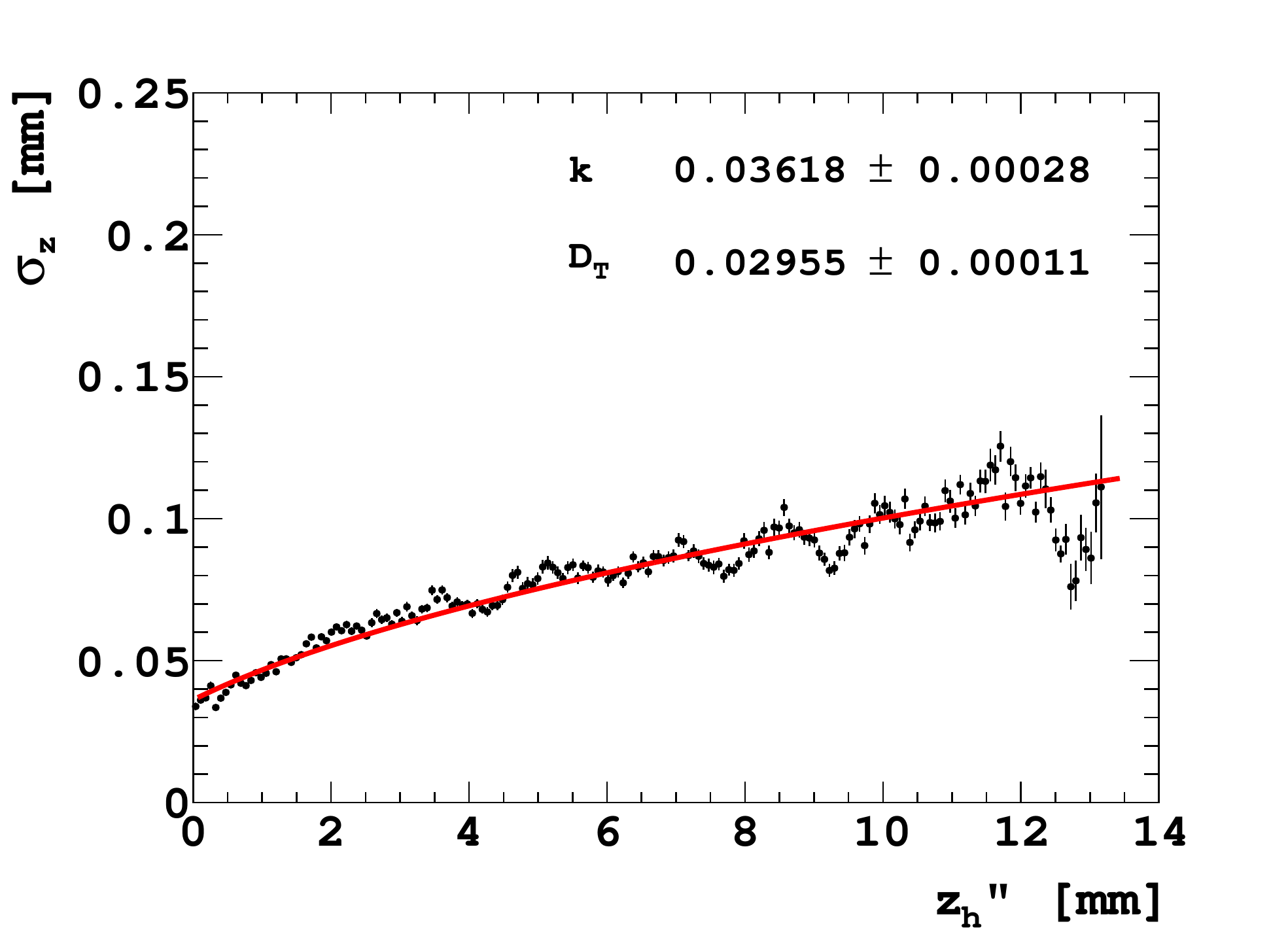}
\caption{The error on the $z$ position as a function of the ionisation height $z^{''}_h$. The data points represent the width of the residuals. The longitudinal diffusion is extracted by fitting the data points with eq.~\ref{eq_fitdiff}.}
\label{fig_diffusion}
\end{figure}

The error on $z$ is plotted against $z^{''}_h$. By fitting the data points with eq.~\ref{eq_fitdiff} we extract the longitudinal diffusion coefficient. Fig.~\ref{fig_diffusion}  shows the result.    

\begin{equation}
\sigma_{z} = \sqrt{k^2 + D^2_L\,z\,\,} .  \label{eq_fitdiff}
\end{equation}

In eq.~\ref{eq_fitdiff}, the offset $k$ includes the errors of the time bin and the timewalk. From the fit we obtain $k = 36.18~\micron$ and $D_L = 29.55~\micron/\sqrt{\mm}$. For CO$_\mathrm{2}$/DME the diffusion coefficient for the transverse and the longitudinal diffusion is approximately the same, \cite{ref_Thesis}. Hence, in the analysis we use $D_L = D_T= 29.55~\micron/\sqrt{\mm}$.

The single hit position resolution in 3D is defined in terms of a voxel. The electron could have been created anywhere within the voxel.  
Tab.~\ref{tab_hitRes} shows the size of the voxel at various ionisation heights. 

\begin{table}[H]
\centering
\caption{The hit position resolution at various ionisation heights.}
\begin{tabular}{rrrr}
\toprule
$z$~($\mm$) & $\sigma_x$~($\micron$) & $\sigma_y$~($\micron$) & $\sigma_z$~($\micron$) \\
\midrule
0.0    & 17.3 & 17.3 & 36.2               \\
1.0    & 34.3 & 34.3 & 46.7               \\
2.0    & 45.2 & 45.2 & 55.3               \\ 
5.0    & 68.3 & 68.3 & 75.4               \\ 
10.0    & 95.0 & 95.0 & 100.2               \\ 
\bottomrule 
\end{tabular}   
\label{tab_hitRes}
\end{table}

\section{Results}
\label{lab_Results}

\subsection{Timewalk correction}

\begin{figure}[!h]
\centering
\includegraphics[width=.9\linewidth]{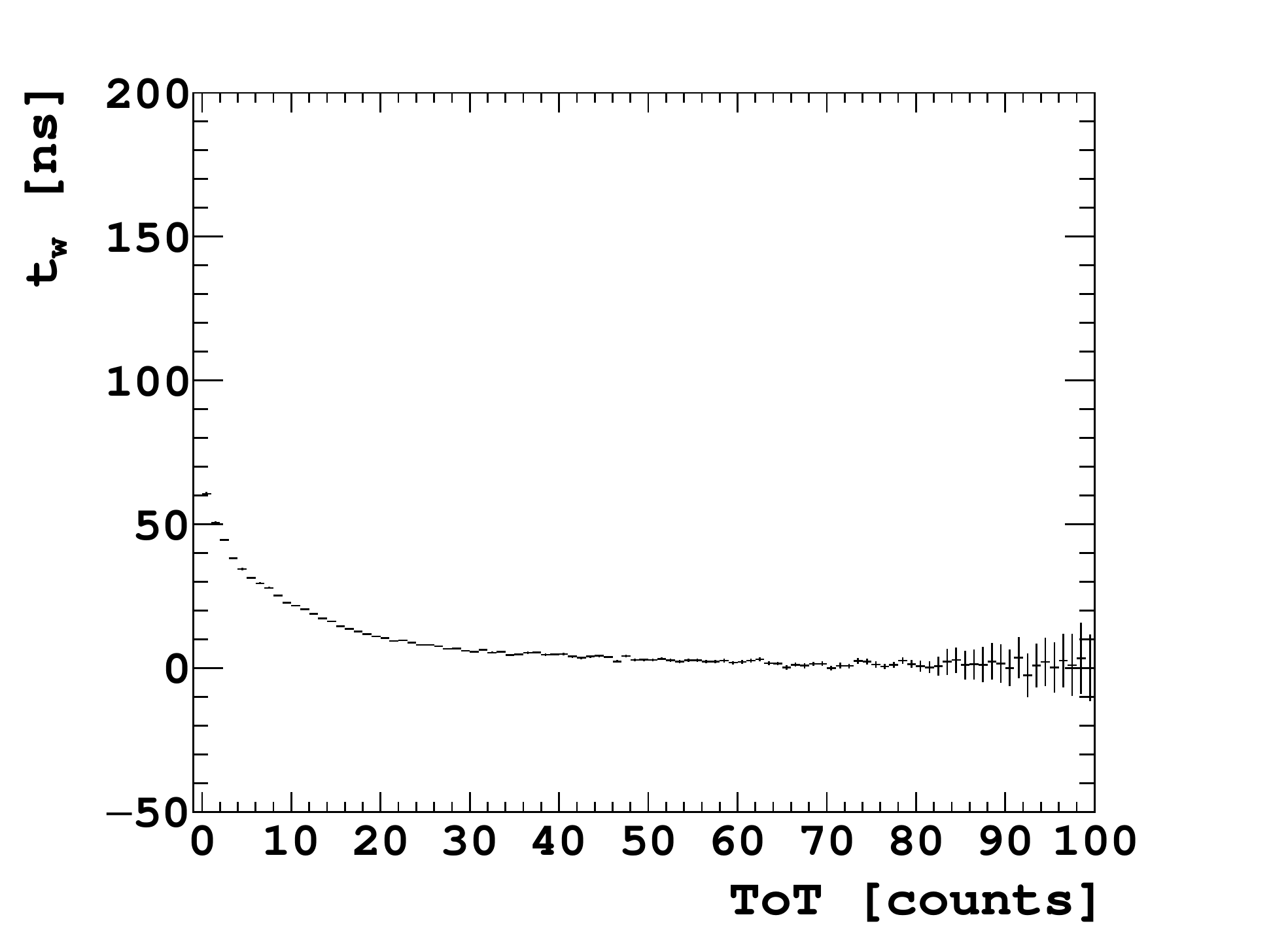}
\caption{The timewalk dependence on the ToT. Each point represents the mean of the gaussian fit of each ToT slice.}
\label{fig_twcorr} 
\end{figure}

\begin{figure}[!h]
\centering
\includegraphics[width=.9\linewidth]{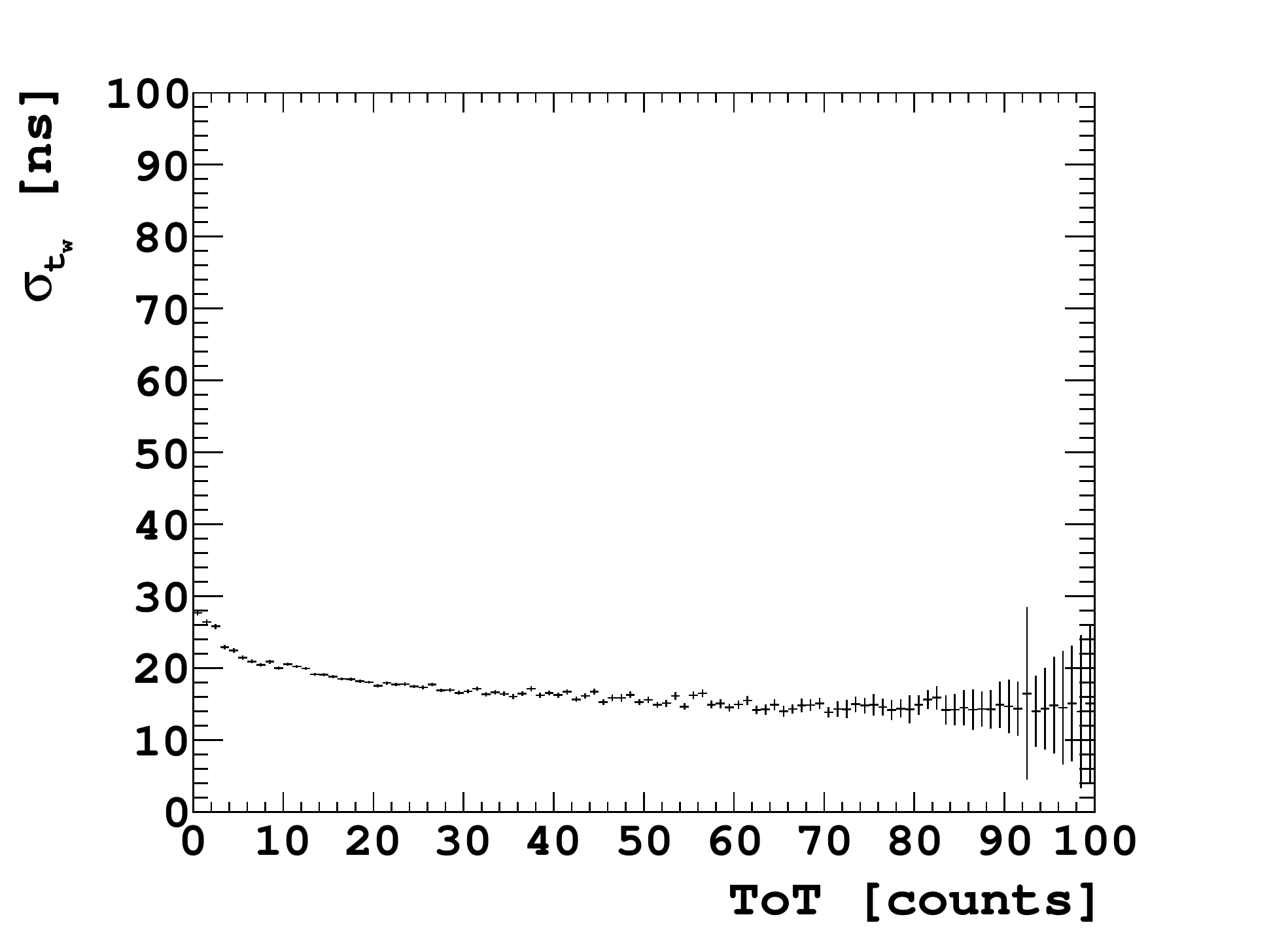}
\caption{The width  of each gaussian against the ToT that is used to extract the corresponding timewalk error.}
\label{fig_twerr} 
\end{figure}

Fig.~\ref{fig_twcorr} shows the fitted mean i.e. the timewalk for each slice of ToT while Fig.~\ref{fig_twerr} shows the fitted width for each ToT slice which we use to extract the contribution of timewalk on the $z^{''}_h$ error. 

The correction is largest at small values of ToT and asymptotic to zero at high values. In the analysis we reject hits with a ToT value lower than five  (ToT~$<5$), the timewalk range is up to 30~$\ns$. The detector during the testbeam was flushed with a CO$_\mathrm{2}$/DME (50/50) gas mixture. For the detector settings used during the testbeam the drift velocity is about 4.5-5.0~$\micron\per\ns$. This results to a correction up  to 135-150~$\micron$, which in the end improves the intrinsic $z$ position resolution.       

The correction is in agreement with the design expectations, \cite{ref_DeGaspari}. At the time when the analysis was complete, this was the first timewalk correction for the Timepix3 chip based on real data from a particle beam, \cite{ref_Thesis}. Previous attempts to estimate timewalk and correct for it were based on test-pulse data.

\subsection{Track position \& angular resolution}

The straight line fit is performed with York's method to the track projections at the $xz$ and the $yz$ plane. The fitting routine returns slope and intercept  for both planes, $(b_x, m_x)$ and $(b_y, m_y)$,  their covariance matrix and the $x^2$, as described in \cite{ref_Gossipo2,ref_WilcoThesis}. The fit is performed at the centre of gravity of the track where the covariance matrix is diagonal. The positional and angular errors are then obtained from the corresponding covariance matrix elements. The average error on the position for the $xz$ and the $yz$ planes is about 12.6~$\micron$ and 11.5~$\micron$ respectively, similar to what has been observed in \cite{ref_Gossipo2}. For the angular error the average error is about 4.4~$\milli\radian$ and 4.3~$\milli\radian$. This is a factor 10 improvement thanks to the larger drift gap and the time-resolution of the Timepix3 chip. 

\begin{figure}[!h]
\centering
\includegraphics[width=0.9\linewidth]{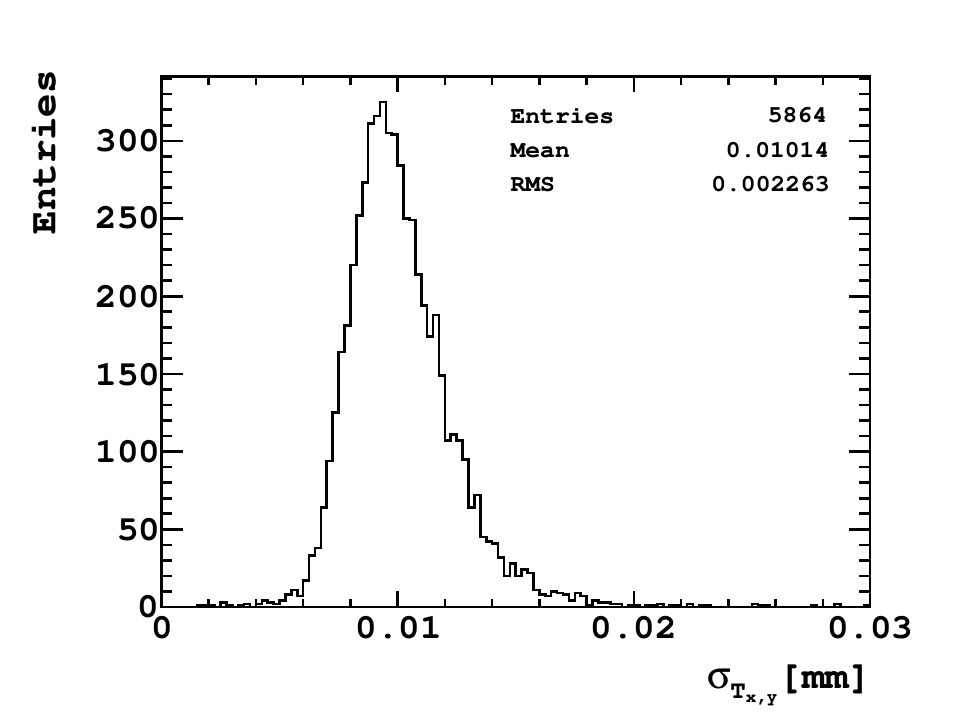}
\caption{The fundamental track position resolution for $x$,$y$. The resolution is exactly the same due to the symmetry of the detector.}
\label{fig_posres_xy}
\end{figure}

By using the positional and angular errors obtained, the set of the fundamental track-resolution parameters of the detector can be derived as in~\cite{ref_Gossipo2}. Due to the symmetry of the detector for $x$ and $y$ direction the track resolution is expected to be the same. Fig.~\ref{fig_posres_xy} shows the result. 

\begin{figure}[!htp]
\centering
\includegraphics[width=0.9\linewidth]{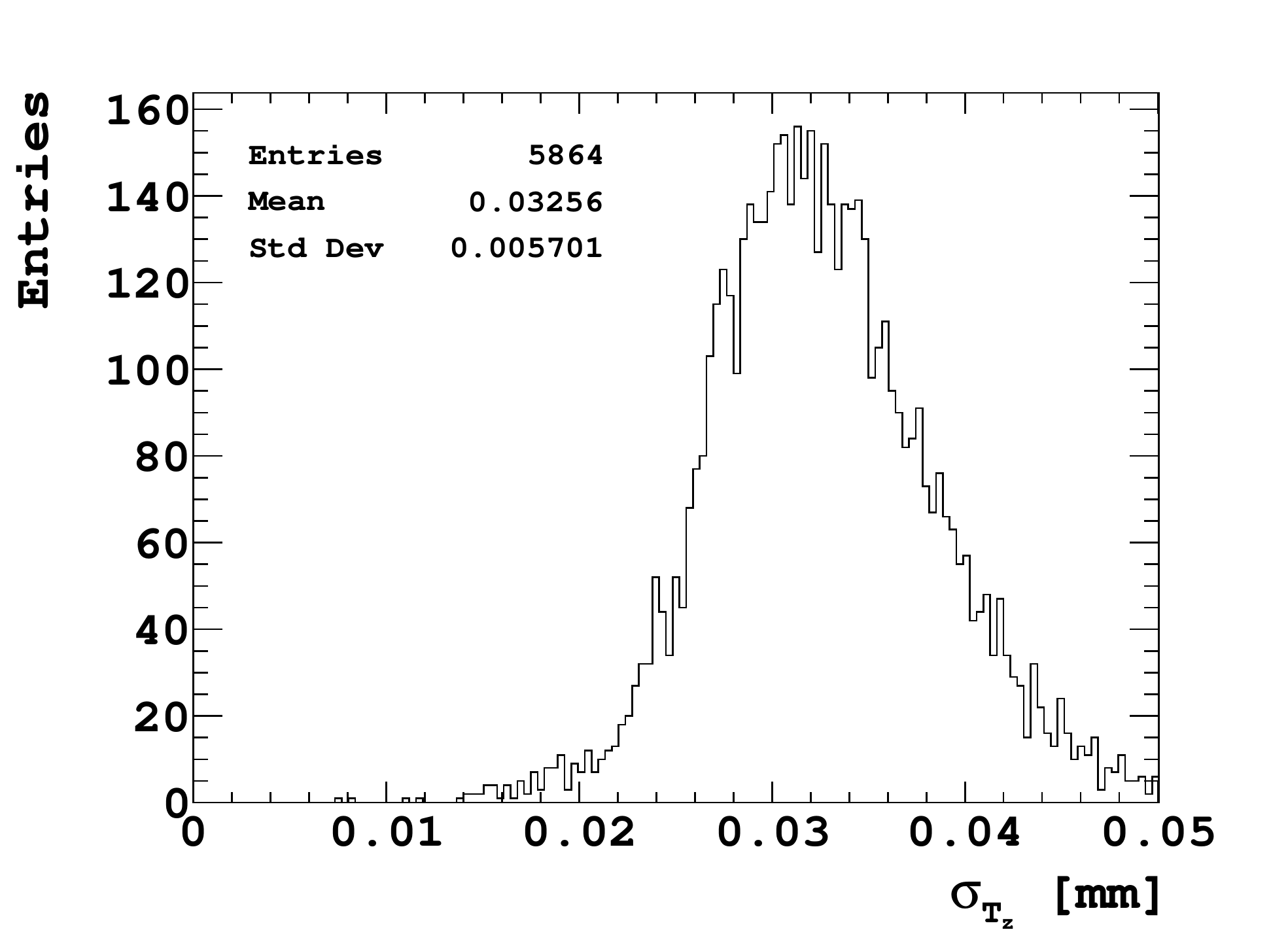}\\
\includegraphics[width=0.9\linewidth]{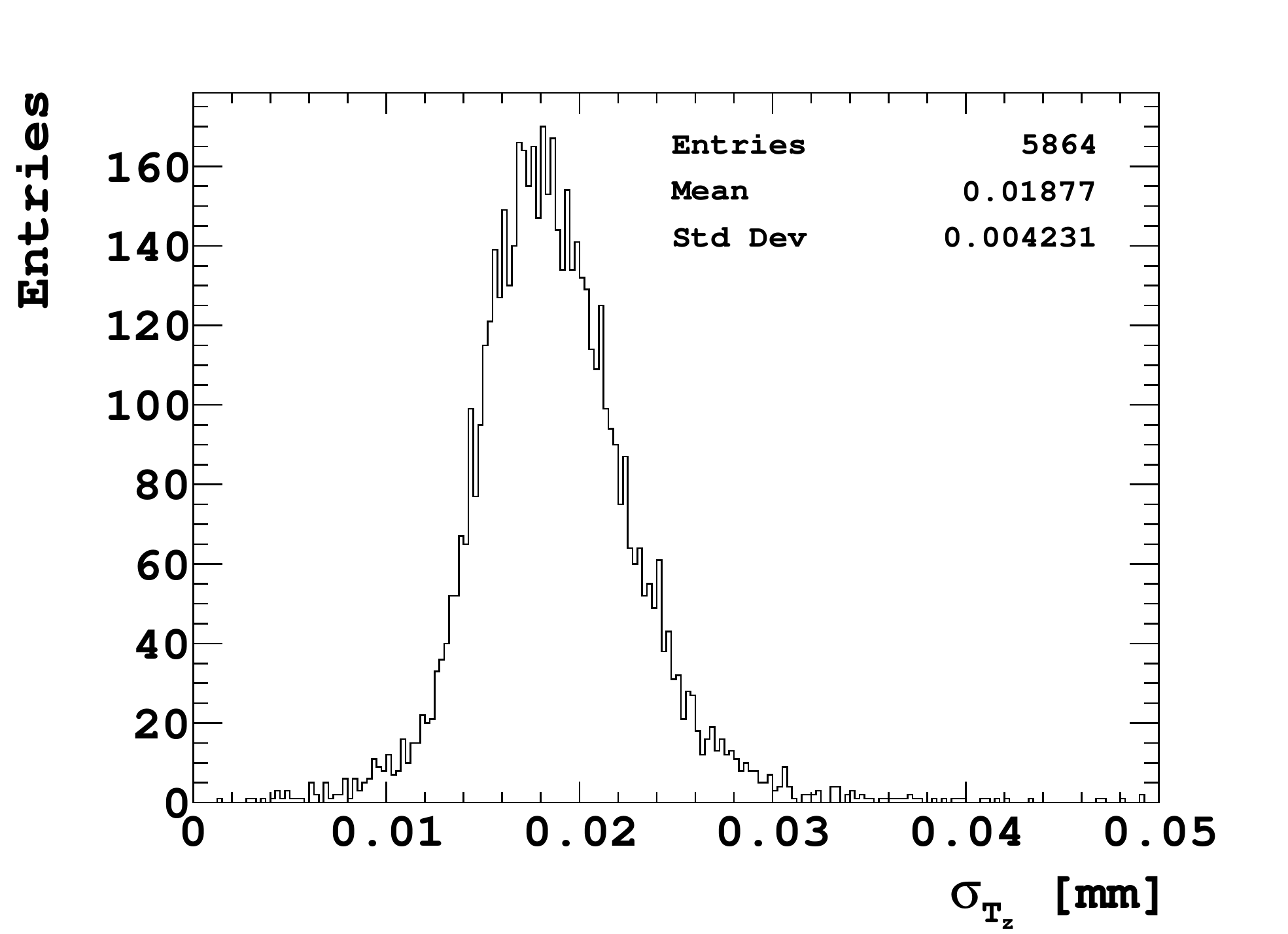}
\caption{The fundamental track position resolution along the drift direction before the timewalk correction on the top and after the timewalk correction on the bottom.}
\label{fig_posres_z}
\end{figure}

The fundamental resolution for $x$ and $y$ is about 10~$\micron$ which is similar to what has been observed before in \cite{ref_Gossipo2}. However more intriguing is the result for the resolution along the drift direction. Fig.~\ref{fig_posres_z} shows the result obtained without the timewalk correction above and after the correction below. Before the correction the resolution averages to 32~$\micron$. Thanks to the timewalk correction the fundamental resolution for $z$ drops to 19~$\micron$. The measured fundamental set of track resolutions (described in \cite{ref_Gossipo2}) is given by eq.~\ref{eq_setReso}: 

\begin{equation}
	(\sigma_{T_x}, \sigma_{T_y}, \sigma_{T_z}) = (10, 10, 19)~\micron. \label{eq_setReso}
\end{equation}

Thanks to the Timepix3 chip which enables the timewalk correction, the improvement with respect to detectors based on Timepix and Gossipo2 chips is significant. 


\section{Conclusions}

We have developed a gaseous pixel detector by mounting a micromegas amplification grid on top of the Timepix3 chip. The detector was tested in a particle beam at CERN. During the testbeam period we have successfully demonstrated the capabilities of the chip and the readout system by recording all tracks. 

The Timepix3 chip thanks to its fast front-end and the simultaneous measurement of ToA and ToT allows the correction for the timewalk effect. We present the timewalk correction for the Timepix3 chip obtained  with the detector operating at the minimum threshold $\sim$~500~$e^-$. The correction is feasible thanks to the telescope which provides a very precise track as an external reference. This correction is consistent with the expectations of the chip designers and could be useful in applications where there is no external reference.  

By applying the timewalk correction on the data, we achieve significant improvement to the position resolution along the drift direction with respect to \cite{ref_Gossipo2}. We present an in-plane position resolution of 10~$\micron$ and 19~$\micron$ for the drift direction. We also present an improved angular resolution with respect to previous works, about 4.5~$\milli\radian$ thanks to much wider pixel plane. Table~\ref{tab_hitRes} shows the single hit position resolution for various drift heights. Although we were hindered by the mismatch of the grid to the cell pitch of the pixel chip and imperfections of the drift field, the data still enabled us to deduce the statistical error in timewalk and diffusion.

Without these imperfections only the timewalk correction would be needed. When using a dedicated integrated grid the number of hits would be about doubled, improving position and angular track resolution by $1/\sqrt{2}$ as it was demonstrated in  \cite{ref_Gossipo2}.

\section*{Acknowledgements}

This work was funded by the Foundation for Fundamental Research on Matter (FOM), which is part of the Netherlands Organisation for Scientific Research (NWO). We would like to thank J. Rovekamp (Nikhef) who performed the wire-bonding. We also thank the LHCb VELO upgrade group for giving us the possibility to make use of the silicon telescope and the CERN/SPS people for the hosting and the beam.

\bibliographystyle{plain} 

\begin{thebibliography}{9}

\bibitem{ref_Gossipo2} S. Tsigaridas, et at., \href{http://dx.doi.org/10.1016/j.nima.2015.06.005}{Precision tracking with a single gaseous pixel detector}, Nuclear Instruments and Methods in Physics Research Section A 795 (2015) 309 - 317.

\bibitem{ref_Timepix3} T. Poikela, et al., \href{http://stacks.iop.org/1748-0221/9/i=05/a=C05013}{Timepix3: a 65K channel hybrid pixel readout chip with simultaneous ToA/ToT and sparse readout}  Journal of Instrumentation 9 (2014) C05013.

\bibitem{ref_etching} A. Delbart and et al., \href{http://dx.doi.org/10.1016/S0168-9002(00)01175-X}{New developments of Micromegas detector}, Nuclear Instruments and Methods in Physics Research Section A 461 (1-3) (2001) 84 – 87.

\bibitem{ref_spidr} J. Visser and et al., \href{http://stacks.iop.org/1748-0221/10/i=12/a=C12028}{SPIDR: a read-out system for Medipix3 \& Timepix3, Journal of Instrumentation}, 10 (12) (2015) C12028.

\bibitem{ref_Thesis} S. Tsigaridas, New Generation GridPix: Development and Characterisation of pixelated gaseous detectors based on the Timepix3 chip, Ph.D. Dissertation, University of Amsterdam, Amsterdam, (2017), \url{http://hdl.handle.net/11245.1/5592b0c6-1559-4a93-af5b-0bb80d520bf4}.

\bibitem{ref_DeGaspari} M. De Gaspari and et al.,\href{http://stacks.iop.org/1748-0221/9/i=01/a=C01037}{Design of the analog front-end for the Timepix3 and Smallpix hybrid pixel detectors in 130 nm CMOS technology} Journal of Instrumentation 9 (01) (2014) C01037.

\bibitem{ref_WilcoThesis} W. Koppert, GridPix: Development and Characterisation of a Gaseous Tracking Detector, Ph.D. Dissertation, University of Amsterdam, Amsterdam, (2015), \url{http://hdl.handle.net/11245/1.437811}.

\end{thebibliography}

\end{document}